\documentclass{aa}
\usepackage{savesym}
\usepackage{amsmath}
\savesymbol{iint}
\usepackage{txfonts}
\restoresymbol{TXF}{iint}
\usepackage{graphics,graphicx}
\usepackage{amssymb}
\usepackage{color}
\usepackage{breakurl}
\usepackage{array}
\usepackage{rotating}
\usepackage{xcolor}
\usepackage{natbib}
\usepackage{float}
\usepackage{flafter}
\usepackage{booktabs}
\usepackage{afterpage}
\usepackage{lipsum}
\usepackage{longtable,pdflscape}
\bibpunct{(}{)}{;}{a}{}{,}

\newcommand{\SNR}{{SNR~J0533$-$7202}}

\newcommand{\xmm}{{\it XMM-Newton}}
\newcommand{\rosat}{{\it ROSAT}}

\newcommand{\spitzer}{{\it Spitzer}}
\newcommand{\source}{{XMMU~J053348.2$-$720233}}

\newcommand{\vpshock}{{\tt vpshock}}
\newcommand{\vsedov}{{\tt vsedov}}
\newcommand{\vphabs}{{\tt vphabs}}

\newcommand{\phabs}{{\tt phabs}}

\begin{document}

\title{{\it XMM-Newton} observation of \SNR\ in the \\ Large Magellanic Cloud\thanks{Based on observations obtained with \xmm, an ESA science mission with instruments and contributions directly funded by ESA Member States and NASA}}



\author{P.~J.~Kavanagh \inst{1} \and M.~Sasaki \inst{1} \and E.~T.~Whelan\inst{1} \and P.~Maggi \inst{2}\thanks{Now at Laboratoire AIM, CEA-IRFU/CNRS/Universit\'e Paris Diderot, Service d'Astrophysique, CEA Saclay, F-91191 Gif sur Yvette Cedex, France} \and F.~Haberl \inst{2} \and L.~M.~Bozzetto \inst{3} \and \\ M.~D.~Filipovi\'c \inst{3} \and E.~J.~Crawford \inst{3}}

 

\institute{Institut f\"{u}r Astronomie und Astrophysik, Kepler Center for Astro and Particle Physics, Eberhard Karls Universit\"{a}t T\"{u}bingen, Sand~1, T\"{u}bingen D-72076, Germany\\ \email{kavanagh@astro.uni-tuebingen.de}
\and Max-Planck-Institut f\"{u}r extraterrestrische Physik, Giessenbachstra\ss e, D-85748 Garching, Germany
\and University of Western Sydney, Locked Bag 1791, Penrith, NSW 2751, Australia
}

\date{Received ?? / Accepted ??}

\abstract{}{We present an X-ray study of the supernova remnant \SNR\ in the Large Magellanic Cloud (LMC) and determine its physical characteristics based on its X-ray emission.}{We observed \SNR\ with \xmm\ (background flare-filtered exposure times of 18 ks EPIC-pn and 31 ks EPIC-MOS1, EPIC-MOS2). We produced X-ray images of the supernova remnant, performed an X-ray spectral analysis, and compared the results to multi-wavelength studies.}{The distribution of X-ray emission is highly non-uniform, with the south-west region much brighter than the north-east. The detected X-ray emission is correlated with the radio emission from the remnant. We determine that this morphology is most likely due to the supernova remnant expanding into a non-uniform ambient medium and not an absorption effect. We estimate the remnant size to be $53.9~(\pm3.4)\times43.6~(\pm3.4)$~pc, with the major axis rotated $\sim64^{\circ}$ east of north. We find no spectral signatures of ejecta emission and infer that the X-ray plasma is dominated by swept up interstellar medium. Using the spectral fit results and the Sedov self-similar solution, we estimate the age of \SNR\ to be $\sim17-27$~kyr, with an initial explosion energy of $(0.09-0.83)\times10^{51}$~erg. We detected an X-ray source located near the centre of the remnant, namely \source. The source type could not be conclusively determined due to the lack of a multi-wavelength counterpart and low X-ray counts. We found that it is likely either a background active galactic nucleus or a low-mass X-ray binary in the LMC.}{We detected bright thermal X-ray emission from \SNR\ and determined that the remnant is in the Sedov phase of its evolution. The lack of ejecta emission prohibits us from typing the remnant with the X-ray data. Therefore, the likely Type~Ia classification based on the local stellar population and star formation history reported in the literature cannot be improved upon.}

\titlerunning{{\it XMM-Newton} observation of \SNR}
\maketitle 

\section{Introduction}
The study of supernova remnants (SNRs) is important in many fields of astrophysics. SNRs distribute the heavy elements formed inside their stellar progenitors and via explosive nucleosynthesis in their supernovae (SNe) into the interstellar medium (ISM) of their host galaxies. The kinetic energy they impart drives the physical evolution of the ISM and cosmic rays can be accelerated in their fast moving shocks. Studies of these objects in the Milky Way are heavily affected by many factors including uncertain distance estimates and high foreground absorption. This is especially problematic for the study of evolved SNRs, whose low temperature plasmas are particularly susceptible to foreground absorption. To avoid these biases in studies of SNRs, we must look to the nearby galaxies of the Local Group.

\par The \object{Large Magellanic Cloud} (\object{LMC}) is the quintessential target for the study of a population of SNRs. At a distance of 50~kpc \citep{diBen2008} it is sufficiently close that its stellar population and diffuse structure is resolved in most wavelength regimes. The LMC is almost face-on \citep[inclination angle of 30--40$^\circ$,][]{vanderMar2001,Nikolaev2004} making the entire population of SNRs available for study. The modest extinction in the line of sight (average Galactic foreground $N_{\rm{H}} \approx 7 \times 10^{20}$~cm$^{-2}$) means optical and X-ray observations of SNRs are only slightly affected by foreground absorption, whereas its location in one of the coldest parts of the radio sky \citep{Haynes1991} allows for improved radio observations without interference from Galactic emission. These favourable conditions have allowed a substantial population of SNRs to be identified in this galaxy. \citet{Badenes2010} compiled a catalogue of 54 SNRs in the LMC. However, this number has been steadily increasing in recent years because of improved observation sensitivity and/or multi-wavelength confirmation of candidates. Recent additions include \citet{Kavanagh2013}, \citet{Bozzetto2014}, \citet{Maggi2014}, and \citet{Kavanagh2015}.


\begin{figure*}
\begin{center}
\resizebox{\hsize}{!}{\includegraphics[trim= 0cm 0cm 0cm 0cm, clip=true, angle=0]{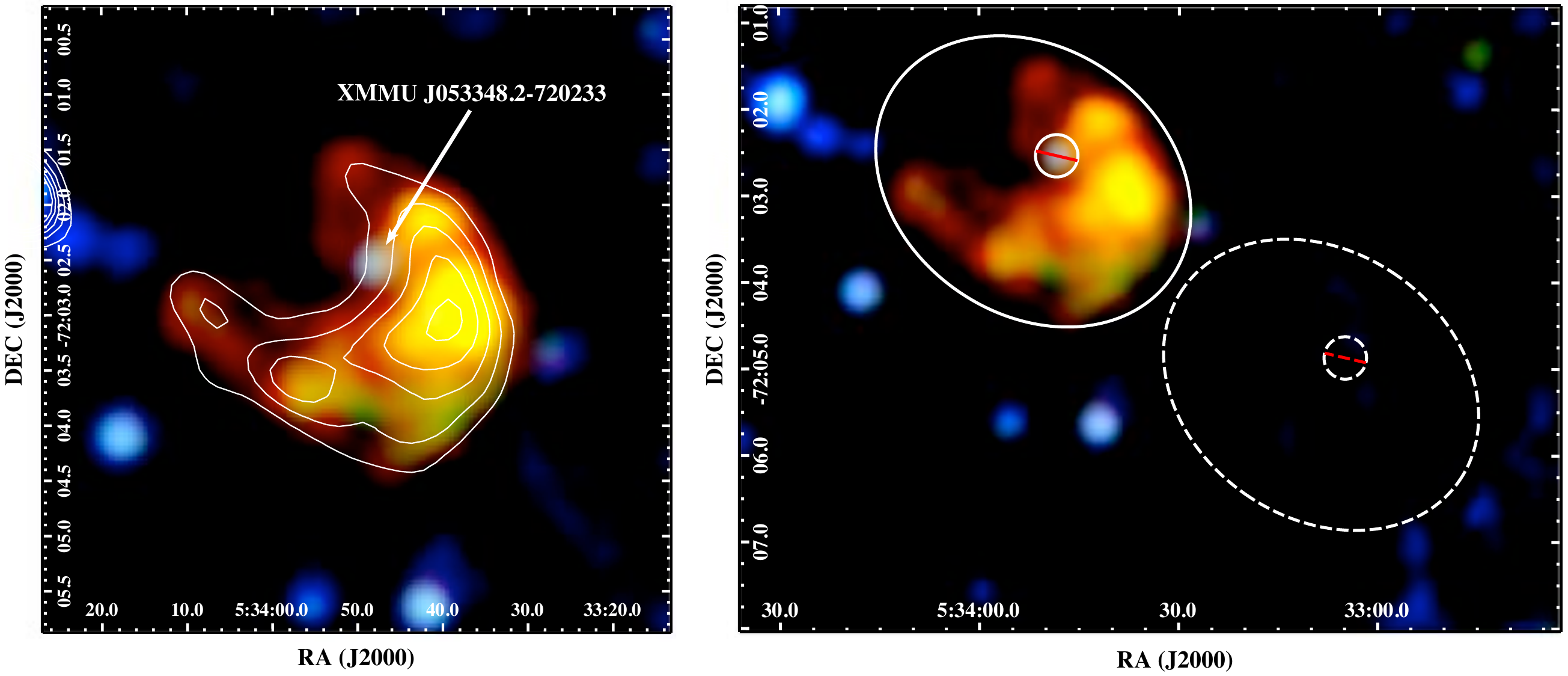}}
\caption{{\bf Left:} \xmm\ EPIC image of \SNR\ in false colour with RGB corresponding to 0.3--0.7~keV, 0.7--1.1~keV, and 1.1--4.2~keV. The image is overlaid with 6~cm (5500~MHz) contours with levels corresponding to 3, 6, 9, 12, and 15$\sigma$ (from BF13, Fig.~1). The location of the point source \source\ is shown. {\bf Right:} Same as $Left$ but with a zoomed out view showing the SNR and background spectral extraction regions.}
\label{snrs_im}
\end{center}
\end{figure*}

\par \citet[][hereafter BF13]{Bozzetto2013a} used radio observations by the Australia Telescope Compact Array (ATCA) to confirm the SNR nature of an unclassified object in the Magellanic Cloud Surveys of \citet{Filipovic1995,Filipovic1998a,Filipovic1998b}, assigning the identifier \object{SNR~J0533$-$7202}. The SNR exhibits a horseshoe morphology with an estimated size of $37(\pm1)\times28(\pm1)$~pc. This differs from the dimensions determined in this work and the reasons for this are discussed in Section~\ref{mwm}. BF13 report no significant optical line emission, with only very tentative H$\alpha$ emission detected in the Magellanic Cloud Emission Line Survey \citep[MCELS,][]{Smith2006} images. Similarly, these authors find no infrared (IR) emission associated with \SNR\ in the \spitzer\ mosaic images of the LMC \citep{Meixner2006}. The \rosat\ source \object{1RXSJ053353.6-720404} \citep{Voges1999} is located at the rim of the radio SNR. However, the source is very faint with only enough count statistics for a position determination. BF13 used data from the Magellanic Cloud Photometric Survey catalogue \citep{Zaritsky2004} and the star formation history map of the LMC \citep{Harris2009} to determine that \SNR\ is associated with an old stellar population, free of recent star formation, and is therefore most likely the result of a Type~Ia explosion.

\par In this paper we report on a follow-up X-ray observation of \SNR\ with \xmm. This observation and data analysis are described in Section~2.  The results from the observation are explained in Section~3. Finally, we summarise our work in Section~4.

\section{Observations and data analysis}
\subsection{X-ray observation}
\textit{XMM-Newton} \citep{Jansen2001} observed \SNR\ on May~9~2013 (Obs. ID 0720440101, PI M. Sasaki). The primary instrument for the observation was the European Photon imaging Camera (EPIC), which consists of a pn CCD \citep{Struder2001} and two MOS CCD \citep{Turner2001} imaging spectrometers. The observational data were reduced using the standard reduction tasks of SAS\footnote{Science Analysis Software, see http://xmm.esac.esa.int/sas/} version 13.5.0, filtering for periods of high particle background. This resulted in flare-filtered exposure times of $\sim18$~ks for the EPIC-pn and $\sim31$~ks for both the EPIC-MOS1 and EPIC-MOS2 detectors.

\subsection{X-ray imaging}
\label{x-ray-imaging}
We produced images and exposure maps in various energy bands from the flare-filtered event lists for each EPIC instrument. We filtered for single and double-pixel events (\texttt{PATTERN} $\lesssim4$) from the EPIC-pn detector, with only single pixel events considered below 0.5 keV to avoid the higher detector noise contribution from the double-pixel events at these energies. All single to quadruple-pixel events (\texttt{PATTERN} $\lesssim12$) were considered for the MOS detectors. 

\par We used three energy bands suited to the analysis of the spectra of SNRs. A soft band from 0.3--0.7 keV includes strong lines from O; a medium band from 0.7--1.1 keV comprises Fe L-shell lines as well as Ne He$\alpha$ and Ly$\alpha$ lines; and a hard band (1.1--4.2 keV) which includes lines from Mg, Si, S, Ca, and Ar.

\par We subtracted the detector background from the images using filter-wheel-closed data (FWC). The contribution of the detector background to each EPIC detector was estimated from the count rates in the corner of the images, which were not exposed to the sky. We then subtracted appropriately-scaled FWC data from the raw images. We merged the EPIC-pn and EPIC-MOS images into combined EPIC images and performed adaptive smoothing of each using an adaptive template determined from the combined energy band (0.3--4.2~keV) EPIC image. The sizes of Gaussian kernels were computed at each position in order to reach a signal-to-noise ratio of five, setting the minimum full width at half maximum of the kernels to $\sim8\arcsec$. In the end the smoothed images were divided by the corresponding vignetted exposure maps. Finally, we produced a three-colour image of \SNR\, which is shown in Fig. \ref{snrs_im}-left.

\subsection{Point sources}
\label{src_proc}
We searched for point sources associated with \SNR\ using the SAS task \texttt{edetect\_chain}. Images were extracted from each of the EPIC instruments in the standard energy bands 0.2--0.5~keV, 0.5--1~keV, 1--2~keV, 2--4.5~keV, and 4.5--12~keV, with the same pattern filtering criteria outlined in Section \ref{x-ray-imaging}

\par For an adopted minimum-detection likelihood\footnote{The detection likelihood $L$ is defined by $L = - \rm{ln}$~$p$, where $p$ is the probability that a Poissonian fluctuation of the background is detected as a spurious source.} of 10, we detected several sources in and around \SNR. However, most of these were false detections due to the bright extended emission regions of the remnant. After visual inspection and false source exclusion, only one was found to reside within the extent of \SNR. This source is located close to the remnant centre, as indicated in Fig.~\ref{snrs_im}-left.

We corrected the X-ray positions of the \xmm\ sources by calculating the offsets of those stars in the field-of-view (FOV) with optical and infrared (IR) counterparts using the USNO-B1 and 2MASS catalogues of \citet{Monet2003} and \citet{Cutri2003}, respectively. The offset between the X-ray positions and the optical/IR positions corrected for proper motion is $\Delta\rm{RA} = -0.03 \pm 0.54$~arcsec and $\Delta\rm{Dec} = 0.54 \pm 0.80$~arcsec. Since both offsets are not statistically significant, we do not apply an astrometric correction to the X-ray source positions. Therefore, we determined J2000 coordinates of RA~=~05$^{\rm{h}}$33$^{\rm{m}}$48.29$^{\rm{s}}$ and Dec~=~$-72$$^{\rm{d}}$02$^{\rm{m}}$33.0$^{\rm{s}}$ for the source near the centre of \SNR\ and assign the identifier \object{XMMU~J053348.2$-$720233}. The analysis and discussion of \source\ and its relation to \SNR\ is given in Section~\ref{inner_source}.

\subsection{X-ray spectral analysis}
\label{spec_analysis}
For the spectral analysis, we made use of the EPIC-pn and EPIC-MOS data. Although less sensitive, the EPIC-MOS spectral resolution is slightly better than the EPIC-pn, and therefore helps to constrain the parameters of the spectral models. Before proceeding with the spectral extraction we generated vignetting-weighted event lists for each EPIC instrument to correct for the effective area variation across our extended source using the SAS task \texttt{evigweight}. Spectra were extracted from elliptical regions encompassing the X-ray extent of the SNR. Backgrounds were extracted from a source and diffuse emission free region immediately to the south-west (see Fig.~\ref{snrs_im}-right). Detected point sources in the source and background regions were excluded. All spectra were rebinned so that each bin contained a minimum of 30 counts to allow the use of the $\chi^{2}$ statistic during spectral fitting. The EPIC-pn and EPIC-MOS source and background spectra were fitted simultaneously using XSPEC \citep{Arnaud1996} version 12.8.1 with abundance tables set to those of \citet{Wilms2000}, and photoelectric absorption cross-sections set to those of \citet{Bal1992}.

\subsubsection{X-ray background}
\label{x-ray_background}
Detailed descriptions of the X-ray background constituents and spectral modelling can be found in \citet{Bozzetto2014} and \citet{Maggi2014}. Here we briefly summarise the treatment of the X-ray background in the case of \SNR.

The X-ray background consists of the astrophysical X-ray background (AXB) and particle induced background.
The AXB typically comprises four or fewer components \citep{Snowden2008,Kuntz2010}, namely the unabsorbed thermal emission from the Local Hot Bubble, absorbed cool and hot thermal emission from the Galactic halo, and an absorbed power law representing unresolved background active galactic nuclei (AGN). The spectral properties of the background AGN component were fixed to the well known values of $\Gamma \sim 1.46$ and a normalisation equivalent to 10.5 photons~keV~cm$^{-2}$~s$^{-1}$~sr$^{-1}$ \citep{Chen1997}. The foreground absorbing material comprises both Galactic and LMC components. The foreground Galactic absorption component was fixed at $7.4\times10^{20}$~cm$^{-2}$ based on the \citet{Dickey1990} HI maps, determined using the HEASARC $N_{\rm{H}}$ Tool\footnote{\burl{http://heasarc.gsfc.nasa.gov/cgi-bin/Tools/w3nh/w3nh.pl}}, while the foreground LMC absorption component, with abundances set to those of the LMC, was allowed to vary. We note that in all the forthcoming fits, the normalisation of the Local Hot Bubble component consistently tended to 0, with only upper limits to its contribution determined. Therefore, since the model is not formally required, we excluded this component from the spectral fits.

The particle-induced background of the EPIC consists of the quiescent particle background (QPB), instrumental fluorescence lines, electronic read-out noise, and residual soft proton (SP) contamination. To determine the contribution of these components we made use of vignetting corrected FWC data. We extracted FWC spectra from the same detector regions as the observational source and background spectra. The EPIC-pn FWC spectra were fitted with the empirical model developed by \citet{SturmPhD} and the EPIC-MOS FWC spectra with an adapted form of this model. Since these spectral components are not subject to the instrumental response, we used a diagonal response in XSPEC. The resulting best-fit model was included and frozen in the fits to the observational spectra, with only the widths and normalisations of the fluorescence lines allowed to vary. We also included a multiplicative constant to normalise the continuum to the observational spectra using the high energy tail ($E>4$~keV) where the QPB component dominates. The residual SP contamination was fitted by a power law not convolved with the instrumental response \citep{Kuntz2008}. 

\subsubsection{Source emission}
\par
\begin{figure}[!t]
\begin{center}
\resizebox{\hsize}{!}{\includegraphics[trim= 0cm 0cm 0cm 0cm, clip=true, angle=0]{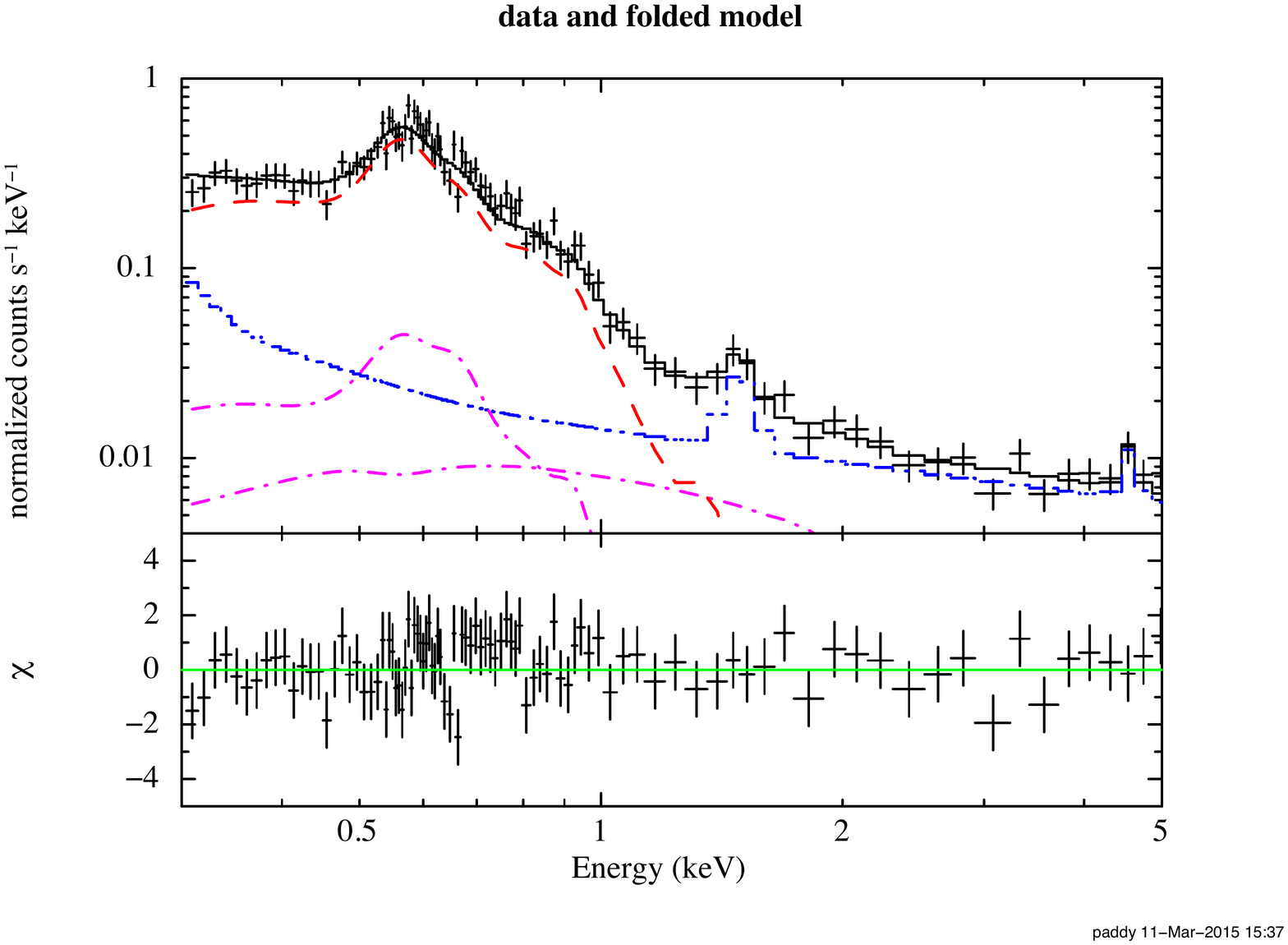}}
\caption{\xmm\ EPIC-pn spectrum of \SNR, shown alone for clarity. The best-fit model is given by the solid black line. The red dashed line indicates the \vsedov\ component corresponding to the SNR emission, the magenta dash-dot lines mark the AXB components, the blue dash-dot-dot-dot line shows the combined contributions of the QPB, residual SPs, instrumental fluorescence lines, and electronic noise. The source fit parameters are given in Table~\ref{0533_tab}. 
}
\label{0533-spec}
\end{center}
\end{figure}

 To account for the X-ray emission from \SNR, we included a thermal plasma model in the spectral fits absorbed by foreground Galactic and LMC material. The abundance of the LMC absorption component was fixed to the LMC values \citep[$0.5~\rm{Z}_\sun$,][]{Russell1992}. We made use of a non-equilibrium ionisation (NEI) model in XSPEC, appropriate for SNRs, namely the plane-parallel \texttt{vpshock} model \citep{Borkowski2001}. The \texttt{vpshock} model features a linear distribution of ionisation ages behind the shock which is more realistic than single ionisation age models such as \texttt{vnei}. Given its relatively large size  (see Section~\ref{mwm}), we can assume that the remnant has most likely entered the Sedov phase of its evolution. Therefore, we also fit the spectrum with the Sedov dynamical model, implemented as \texttt{vsedov} in XSPEC \citep{Borkowski2001}. In all of our spectral fits we allowed the abundances of O and Fe to vary, to determine if the abundances of these elements are consistent with the LMC ISM, or are enhanced due to an ejecta contribution. The results of the fits are shown in Table~\ref{0533_tab} with the best-fit \vsedov\ model shown in Fig.~\ref{0533-spec}.

\begin{table}
\caption{Spectral fit results \SNR. See text for description of the models.}
\begin{center}
\label{0533_tab}
\begin{tabular}{llr}
\hline
Component & Parameter & Value\\
\hline
\hline
\multicolumn{3}{c}{\vpshock}\\
\hline
\multicolumn{3}{c}{ }\\
\phabs & $N_{\rm{H,Gal}}$ ($10^{22}$ cm$^{-2}$) &   0.074\tablefootmark{a}  \\
\vphabs & $N_{\rm{H,LMC}}$ ($10^{22}$ cm$^{-2}$) &   0.00 ($<0.03$)\tablefootmark{b}  \\
\multicolumn{3}{c}{ }\\
\vpshock & $kT$ & 0.36 (0.28--0.47)\tablefootmark{b}  \\
 & O ($Z/\rm{Z}_{\sun}$) & 0.39 (0.33--0.45) \\
 & Fe ($Z/\rm{Z}_{\sun}$) & 0.60 (0.45--0.80) \\
 & $\tau_{u}$ ($10^{11}$~s~cm$^{-3}$) & 0.86 (0.37--2.00) \\
 & $EM$ ($10^{57}$ cm$^{-3}$) & 4.2 (2.6--7.5) \\
\multicolumn{3}{c}{ }\\
 & $F_{X}$\tablefootmark{c} ($10^{-13}$ erg~s$^{-1}$~cm$^{-2}$) & 1.9 \\
 & $L_{X}$\tablefootmark{d} ($10^{35}$ erg~s$^{-1}$) & 1.0 \\
\multicolumn{3}{c}{ }\\
Fit statistic & $\chi^{2}_{\nu}$ & 1.13 (312 d.o.f.) \\
\multicolumn{3}{c}{ }\\
\hline
\multicolumn{3}{c}{\vsedov}\\
\hline
\multicolumn{3}{c}{ }\\
\phabs & $N_{\rm{H,Gal}}$ ($10^{22}$ cm$^{-2}$) &   0.074\tablefootmark{a}  \\
\vphabs & $N_{\rm{H,LMC}}$ ($10^{22}$ cm$^{-2}$) &   0.00 ($<0.02$)\tablefootmark{b}  \\
\multicolumn{3}{c}{ }\\
\vsedov & $kT$ & 0.22 (0.17--0.36)\tablefootmark{b}  \\
 & O ($Z/\rm{Z}_{\sun}$) & 0.48 (0.42--0.54) \\
 & Fe ($Z/\rm{Z}_{\sun}$) & 0.67 (0.46--0.88) \\
 & $\tau$ ($10^{11}$~s~cm$^{-3}$) & 3.25 (0.61--25.43) \\
 & $EM$ ($10^{57}$ cm$^{-3}$) & 6.8 (2.7--16.2) \\
\multicolumn{3}{c}{ }\\
 & $F_{X}$\tablefootmark{c} ($10^{-13}$ erg~s$^{-1}$~cm$^{-2}$) & 1.9 \\
 & $L_{X}$\tablefootmark{d} ($10^{35}$ erg~s$^{-1}$) & 1.0 \\
\multicolumn{3}{c}{ }\\
Fit statistic & $\chi^{2}_{\nu}$ & 1.15 (313 d.o.f.) \\
\multicolumn{3}{c}{ }\\
\hline
\end{tabular}
\tablefoot{The numbers in parentheses are the 90\% confidence intervals.
\tablefoottext{a}{Fixed to the Galactic column density from the \citet{Dickey1990} HI maps.}
\tablefoottext{b}{Absorption and thermal component abundances fixed to those of the LMC.}
\tablefoottext{c}{0.3-10~keV absorbed X-ray flux.}
\tablefoottext{d}{0.3-10~keV de-absorbed X-ray luminosity, adopting a distance of 50~kpc to the LMC.}
}
\end{center}
\end{table}%

\section{Results}
\label{results}
\subsection{\SNR\ morphology}
\label{mwm}
\begin{figure}[!t]
\begin{center}
\resizebox{\hsize}{!}{\includegraphics[trim= 0cm 0cm 0cm 0cm, clip=true, angle=0]{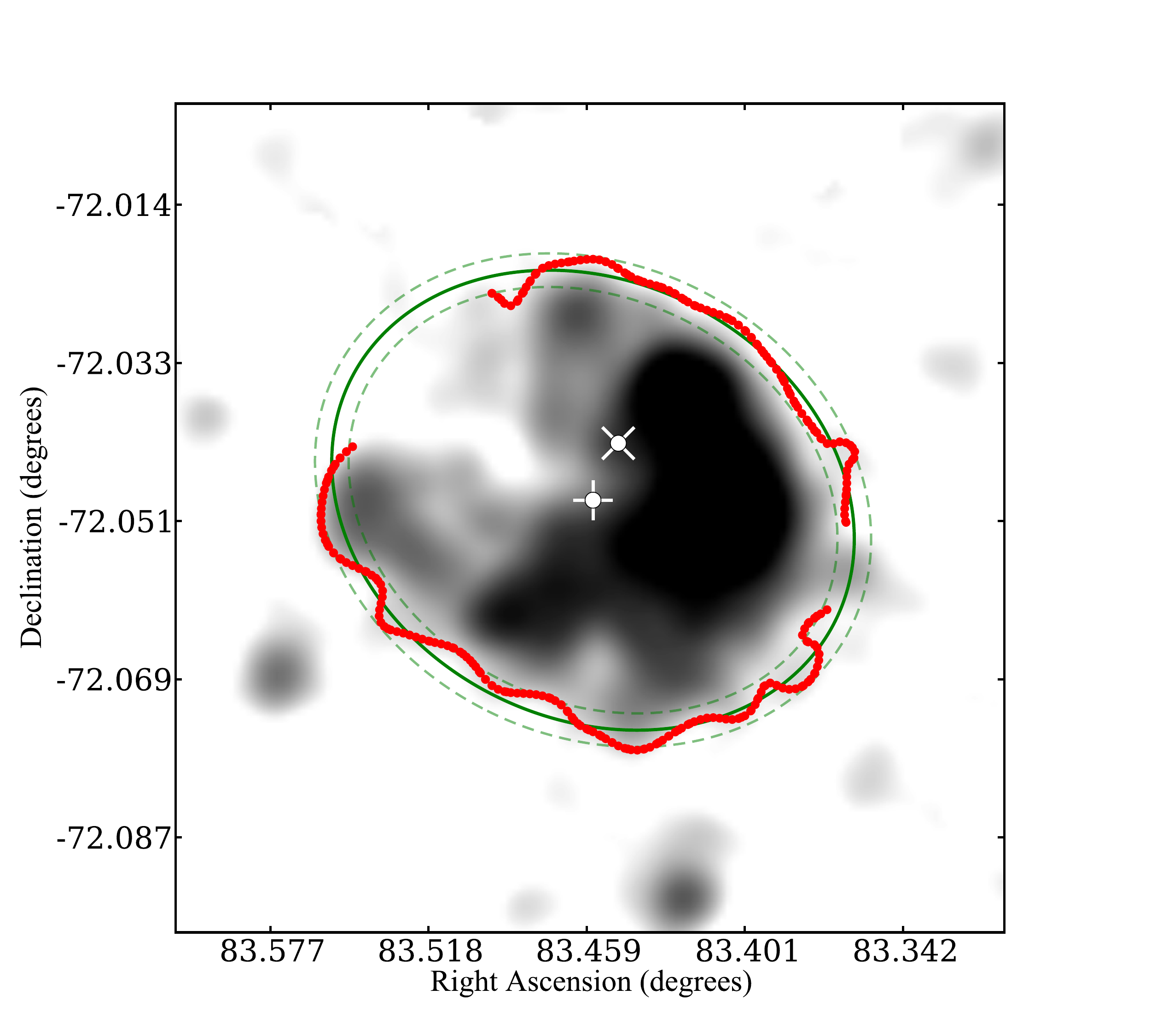}}
\caption{Combined 0.3--0.7~keV EPIC image of \SNR. The red points delineate the contour level corresponding to 3$\sigma$ above the average background surface brightness. The green solid line shows the best-fit ellipse to the contour, with the dashed lines indicating the 1$\sigma$ error on the fit. The white plus-sign marks the best-fit centre of the SNR and the white cross gives the location of \source.}
\label{size}
\end{center}
\end{figure}

The X-ray emission from \SNR\ is highly correlated with the 3~cm and 6~cm radio emission detected in BF13 with the same horseshoe morphology and enhanced emission regions in the south-west (SW, see Fig.~\ref{snrs_im}-left). The X-ray emission is soft with no detection of extended emission in the 1.1--4.2~keV band. Emission in the 0.3--0.7~keV traces the outer edge of the remnant, except for the north-east (NE) limb, pointing to an approximately elliptical shape. To estimate the size of the X-ray remnant we determined the average background surface brightness and corresponding standard deviation ($\sigma$) in the 0.3--0.7~keV band. We then defined the edge of the SNR as regions where the extended emission surface brightness rises to 3$\sigma$ above the average background. We then fitted an ellipse to this contour, excluding the absent limb emission in the NE and an enhancement due to a nearby point source in the SW (see Fig.~\ref{size}). The error on the fit was determined by quantifying the standard deviation of points on the contour from the best-fit ellipse. We determined a best-fit ellipse centred on the J2000 coordinates RA~=~05$^{\rm{h}}$33$^{\rm{m}}$50.19$^{\rm{s}}$ and Dec~=~$-72$$^{\rm{d}}$02$^{\rm{m}}$56.0$^{\rm{s}}$, of size $3\farcm71~(\pm0\farcm23)\times3\farcm00~(\pm0\farcm23)$, corresponding to $53.9~(\pm3.4)\times43.6~(\pm3.4)$~pc at the LMC distance, with the major axis rotated $\sim64^{\circ}$ east of north. The best-fit dimensions and error are shown in Fig.~\ref{size}.

The SNR size in the NW-SE direction (minor axis) is approximately the same as determined in the radio analysis of BF13 (their major axis). However, our NE-SW (major axis) size determined from the X-ray data is much larger than that determined in BF13 (their minor axis). This is most likely due to the size estimation methods employed. BF13 determined the dimensions using a one-dimensional intensity profile along the approximate NE-SW axis, which does not take the elliptical shape of the outer edge of the SNR into account, and only considers the bright SW end.

The asymmetric brightness profile in X-rays could be because of either an ambient density gradient along the NE-SW axis, or because of a higher foreground absorbing column towards the NE of the remnant. To investigate this, we divided the remnant into two halves namely the faint NE and brighter SW regions, and extracted EPIC spectra. We fitted these spectra simultaneously in XSPEC with the thermal plasma models discussed in Section~\ref{spec_analysis}. For the first set of fits, we linked the foreground absorbing column ($N_{\rm{H}}$) towards both regions and only allowed the normalisation of the thermal components to vary to determine if the SW region is intrinsically brighter. For the second set of fits we linked the thermal component parameters of each region, allowing only $N_{\rm{H}}$ to vary to determine if the brightness profile is due to foreground absorption. We found that, in both cases, the foreground absorption values were negligible and only upper limits to the absorption could be determined. These results suggest that the asymmetric brightness profile is intrinsic and due to an ambient density gradient rather higher foreground absorption towards the NE of the remnant.

We also compared the X-ray morphology to the HI maps of \citet{Kim2003} to search for either a density gradient or variations in absorption across the remnant. While the resolution of these HI maps ($1\arcmin$ or 15~pc at the LMC distance) is poorer than \xmm, they do allow for a coarse comparison with the SNR morphology. However, we did not find any conclusive evidence to support either explanation.


\subsection{\SNR\ X-ray emission}
The X-ray spectrum of \SNR\ was found to be well described by either the \vpshock\ or \vsedov\ model with little difference in the fit-statisitic ($\chi^{2}_{\nu}=1.13$ and $\chi^{2}_{\nu}=1.15$, respectively), as shown in Table~\ref{0533_tab}. In each case, the best-fit absorption due to foreground material in the LMC was negligible, with only upper limits determined. The best-fit plasma temperatures for each model agree within the 90\% confidence intervals, and are consistent with other evolved SNRs in the LMC \citep[see][for example]{Bozzetto2014,Maggi2014}. The ionisation ages inferred from the fits are also compatible, though the value is not well constrained in the \vsedov\ model. The best-fit abundances of O and Fe in each model are also in agreement, and are consistent with the LMC abundances \citep{Russell1992}. This suggests that the thermal plasma is dominated by swept-up ISM, with no discernible contribution from shocked ejecta. This justifies our earlier assumption that the SNR is in the Sedov phase of its evolution, and prohibits us from suggesting the type of progenitor supernova responsible for \SNR. 

\begin{table}[t]
\caption{Physical properties of \SNR\ derived from the Sedov model}
\begin{center}
\begin{normalsize}
\label{physical-properties}
\begin{tabular}{ccccc}
\hline
$n_{\rm{0}}$ & $v_{s}$ & $t$ & $M$ & $E_{0}$ \\
 ($10^{-2}$~cm$^{-3}$) & km~s$^{-1}$ & (kyr) & (M$_{\sun}$) & ($10^{51}$ erg) \\
\hline
\hline
2.6 -- 7.0 & 377 -- 549 & 17 -- 27  & 48 -- 162 & 0.09 -- 0.83 \\
\hline
\end{tabular}
\end{normalsize}
\end{center}
\end{table}%

Using the results of the \vsedov\ fit we can estimate physical parameters of \SNR\ using the Sedov dynamical model \citep[see][for examples]{Sasaki2004,Bozzetto2014,Maggi2014}. The X-ray shell of the remnant is delineated by an ellipse (see Section~\ref{mwm}), with semi-major and semi-minor axes of 27.0~pc and 21.8~pc, respectively.  Assuming these are the first and second semi-principal axes of an ellipsoid describing \SNR, and that the third semi-principle axis is in the range 21.8--27.0 pc, we determined the volume ($V$) limits for the remnant and their corresponding effective radii ($R_{\rm{eff}}$) to be $(1.6-2.0)\times10^{60}$ cm$^{3}$ and 23.4--25.1 pc, respectively.\\

\par The best-fit shell X-ray temperature (see Table~\ref{0533_tab}) corresponds to a shock velocity

\begin{equation}
v_{s}=\left(\frac{16kT_{s}}{3\mu}\right)^{0.5},
\end{equation}

\noindent where $kT_{s}$ is the postshock temperature and $\mu$ is the mean mass per free particle. For a fully ionised plasma with LMC abundances, $\mu=0.61\rm{m_{p}}$, resulting in a shock velocity $v_{s}~=~429~(377-549)$~km~s$^{-1}$. The age of the remnant can now be determined from the similarity solution:

\begin{equation}
v_{s}=\frac{2R}{5t},
\end{equation}

\noindent where $R=R_{\rm{eff}}$ and $t$ is the age of the remnant. This gives an age range of $\sim17-27$~kyr. The pre-shock H density ($n_{\rm{H},0}$) in front of the blast wave can be determined from the emission measure ($EM$, see Table \ref{0533_tab}). Evaluating the emission integral over the Sedov solution using the approximation for the radial density distribution of \citet{Kahn1975} gives

\begin{equation}
\int n_{e}n_{\rm{H}}dV=EM=2.07\left(\frac{n_{e}}{n_{\rm{H}}}\right)n_{\rm{H},0}^{2}V,
\end{equation}

\noindent where $n_{\rm{e}}$ and $n_{\rm{H}}$ are electron and hydrogen densities, respectively, and $V$ is the volume \citep[e.g.,][]{Hamilton1983}. Taking $n_{e}/n_{\rm{H}}=1.21$ and the determined range of volumes, this equation yields $n_{\rm{H},0}=(2.4-6.4)\times10^{-2}$~cm$^{-3}$. Since the pre-shock density of nuclei is given as $n_{0}\sim1.1 n_{\rm{H},0}$, it follows that  $n_{0} = (2.6-7.0)\times10^{-2}$~cm$^{-3}$. The initial explosion energy ($E_{0}$) can be determined from the equation:

\begin{equation}
R=\left(\frac{2.02E_{0}t^{2}}{\mu_{n}n_{0}}\right)^{1/5},
\end{equation}

\noindent where $\mu_{n}$ is the mean mass per nucleus, given as $\mu_{n}=1.4m_{p}$. This results in an initial explosion energy in the range $(0.09-0.83)\times10^{51}$~erg. This is slightly lower than the canonical $10^{51}$~erg but consistent with several Type Ia SNRs in the LMC \citep[see][for example]{Hendrick2003,Bozzetto2014}. The swept-up mass contained in the shell is given simply by $M=V\mu_{n}n_{0}$, which is evaluated to ($48-162$)~M$_{\odot}$. This large amount of swept-up material supports our original assumption that \SNR\ is into the Sedov phase of its evolution. All the derived parameters of \SNR\ are summarised in Table \ref{physical-properties}. 

Using the determined values of $t$ and $n_{0}$ we estimated the expected value of the ionisation parameter ($\tau$) in the Sedov spectral fit to the remnant. Given that $\tau$ is the product of the electron density immediately behind the shock front and the remnant age \citep{Borkowski2001}, it can be written $\tau = n_{e}t \approx 4.8n_{\rm{H},0}t$. We determined the expected $\tau$ to be in the range $(0.60-2.53)\times10^{11}$~cm$^{-3}$~s, which is in agreement with the $\tau$ determined in the spectral fits (see Table~\ref{0533_tab}), though the upper bound of the fitted ionisation parameter is poorly constrained.

\subsection{\source}
\label{inner_source}
Our source detection identified the solitary X-ray source \source\ projected inside of \SNR\ (see Section~\ref{src_proc} and Fig.~\ref{snrs_im}-left). Its location near the centre of the remnant (see Fig.~\ref{size}) could indicate that the source is a remnant compact object, pointing to a core-collapse origin for \SNR. This would not be consistent with the Type~Ia scenario proposed by BF13, and therefore the classification of this source has important implications for the typing of the SNR.

We searched for a multi-wavelength counterpart in various catalogues in the literature but none could be identified. As such, the X-ray data alone were used to determine a likely classification for the source. \source\ is embedded in the bright extended emission of \SNR. We determined the source and background (an annulus around the source) extraction regions using the SAS task \texttt{region}, considering only the 1--4.5~keV events to improve the contrast between the source and contaminating background emission from the SNR. Source spectra, background spectra, and ancillary files were then extracted from each of the EPIC instruments and merged using the SAS task \texttt{epicspeccombine}. From the combined EPIC spectrum we determined a net total of $\sim70$ counts, much too low to perform a meaningful spectral analysis. Instead, we performed a hardness ratio (HR) analysis which compares the number of counts in certain energy bands to determine the approximate shape of the spectrum, allowing spectral parameters to be inferred. The HRs are defined as 

\begin{equation}
HR_{i} = \dfrac{R_{i+1} - R_{i}}{R_{i+1}+R_{i}}
\end{equation}

\noindent where $R$ corresponds to the net source counts in a given energy band. For our analysis we defined the energy bands as $i_{1} = 0.3-1$~keV, $i_{2} = 1-2$~keV, and $i_{3} = 2-4.5$~keV, which provides two HR values.

We determined the net source counts in the specified energy bands for \source\ and calculated the hardness ratios with the Bayesian method described in \citet{Park2006}, implemented in the associated BEHR code (ver. 12-12-2013)\footnote{See \burl{http://hea-www.harvard.edu/astrostat/behr/}}. This Bayesian method is much more accurate than the classical methods of HR estimation since it assumes that the detection of counts is a random process described by Poisson statistics. In addition, this method can handle the non-detection of counts in one or more energy bands which would cause classical methods to fail.

\begin{figure}[!t]
\begin{center}
\resizebox{\hsize}{!}{\includegraphics[trim= 0cm 0cm 0cm 0cm, clip=true, angle=0]{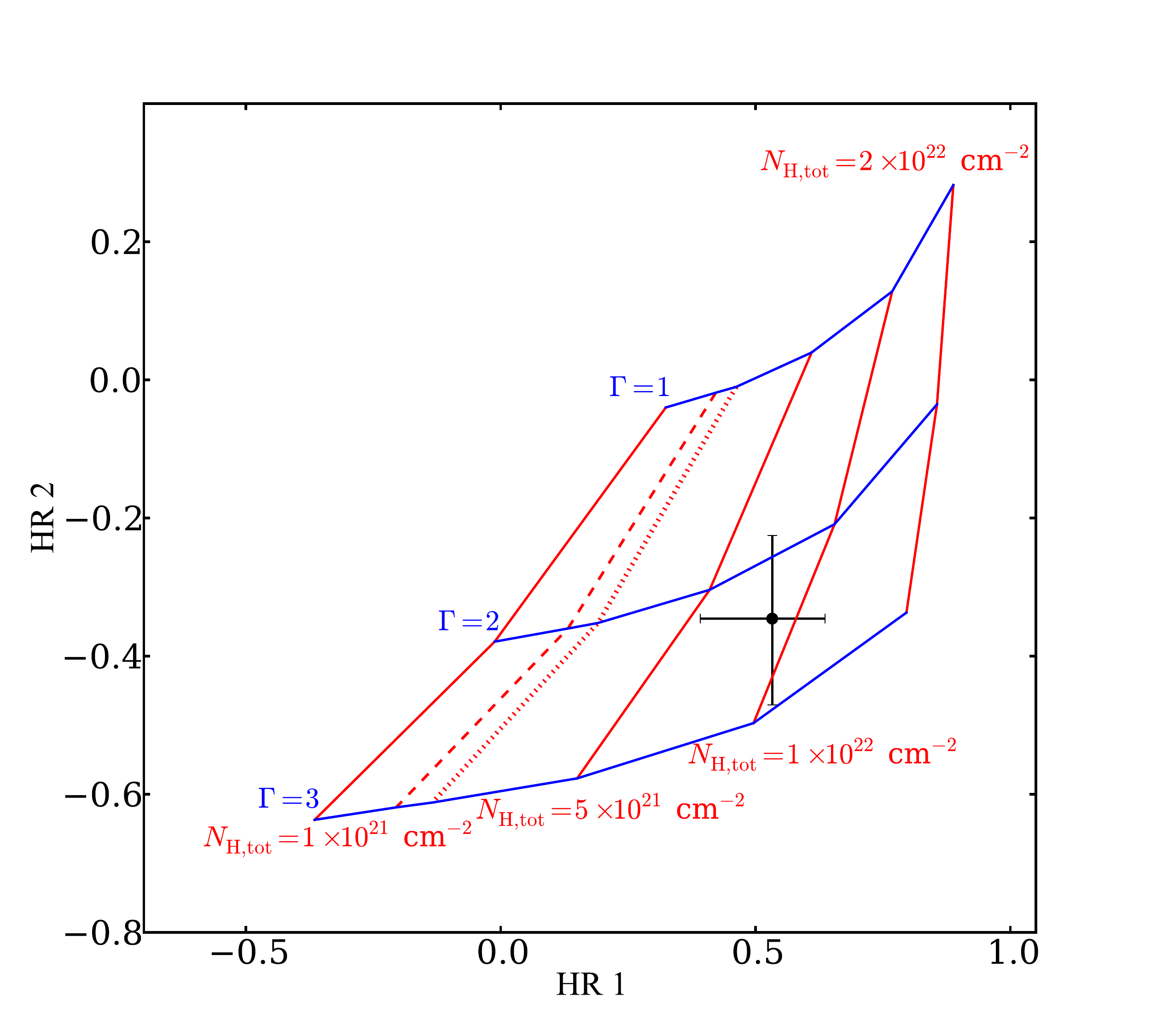}}
\caption{HR plot derived from a grid of simulated spectra, assuming an absorbed power law model, with the HR of \source\ shown in black. See text for a description of the HR simulation and chosen energy bands. The blue lines indicate models on the grid of equal $\Gamma$, whereas the red lines mark the models on the grid of equal equivalent hydrogen column $N_{\rm{H},tot} = N_{\rm{H},Gal} + N_{\rm{H},LMC}$. The dashed red line gives the total $N_{\rm{H}}$ measured value through the Galaxy and the LMC \citep{Kalberla2005} of $2\times10^{21}$~cm$^{-2}$. The dotted red line gives the combined $N_{\rm{H}}$ measured through the Galaxy and the LMC determined by \citet{Dickey1990} and \citet{Kim2003}, respectively, of $2.5\times10^{21}$~cm$^{-2}$.}
\label{hr}
\end{center}
\end{figure}

To determine the likely spectral parameters implied by the HR values of \source, we assumed that the source could be approximately described by a power law model absorbed by material in the Galaxy and LMC (\texttt{phabs*vphabs*pow} in XSPEC). We simulated a set of spectra with fixed Galactic absorption \citep[$N_{\rm{H},Gal}=7.4\times10^{20}$~cm$^{-2}$,][]{Dickey1990}, and varying values of LMC foreground absorption \citep[$N_{\rm{H},LMC}$; with $Z=0.5~\rm{Z}_\sun$,][]{Russell1992} and photon index ($\Gamma$). We produced a grid in HR space and plotted the HR values of \source\ over this, shown in Fig.~\ref{hr}. The source HRs are consistent with that of an absorbed power law with $\Gamma \approx 1.8-3$ and equivalent hydrogen column $N_{\rm{H},tot} = N_{\rm{H},Gal} + N_{\rm{H},LMC} \approx (0.6-1.3)\times10^{22}$~cm$^{-2}$. 

In the absence of a foreground multi-wavelength counterpart, \source\ is likely either located in the LMC or is a background AGN. A background AGN would be subject to foreground absorption by all line-of-sight material in the Galaxy and the LMC, which we estimated in two ways. We first obtained the value of $N_{\rm{H}}$ through the Galaxy and the LMC of $\sim2\times10^{21}$~cm$^{-2}$ from the HI survey of \citet{Kalberla2005}, indicated by the red dashed line in Fig.~\ref{hr}. Alternatively, summing the Galactic $N_{\rm{H}}$ from \citet{Dickey1990} of $7.4\times10^{20}$~cm$^{-2}$ to the LMC $N_{\rm{H}}$ value from \citet{Kim2003} $1.8\times10^{21}$~cm$^{-2}$ gives a combined absorbing column of $\sim2.5\times10^{21}$~cm$^{-2}$, indicated by the red dotted line in Fig.~\ref{hr}. Therefore, the $N_{\rm{H},tot}$ value inferred from the HR analysis suggests that either \source\ is located beyond the LMC and is a background AGN, or is located in the LMC and subject to localised absorption, which could indicate an X-ray binary. In the absence of a high mass stellar population (BF13), we would expect a low-mass X-ray binary (LMXB).  As evident from Fig.~\ref{hr}, these assertions are independent of the chosen HI map. From the likely spectral parameters and the count rate we estimate an indicative unabsorbed 0.3--10~keV flux of the source to be $F_{\rm{X}, 0.3-10~\rm{keV}} \sim 2\times10^{-14}$~erg~cm$^{2}$~s$^{-1}$, corresponding to a luminosity of $L_{\rm{X}, 0.3-10~\rm{keV}} \sim 6\times10^{33}$~erg~s$^{-1}$ if the source is located in the LMC, and could be consistent with a low luminosity LMXB. These values are subject to some uncertainty however given the low count statistics and poorly constrained spectral model. We attempted to perform a timing analysis on the source to search for X-ray pulsations, the period of which could point to either an AGN or LMXB origin. However, the low number of net counts prohibited any conclusive results. In either case, it seems unlikely that the source contains a compact object resulting from a recent core-collapse event.


\section{Summary}
\label{sum}
We presented the analysis of a follow-up \xmm\ observation of \SNR\ in the LMC. The main findings of our analysis can be summarised as follows:

\begin{itemize}
\item We detected bright X-ray emission from \SNR. We estimated the size of the remnant to be $53.9~(\pm3.4)\times43.6~(\pm3.4)$~pc, with the major axis rotated $\sim64^{\circ}$ east of north. The X-ray emission is highly correlated with the radio emission detected by BF13, displaying a highly asymmetric brightness distribution. We found that this is most likely due to an ambient interstellar medium density gradient.

\item The X-ray spectrum of the remnant is thermal in origin and well described by both the non-equilibrium ionisation \vpshock\ model and the Sedov dynamical model. The abundance patterns determined from the spectral fits are consistent with swept-up interstellar medium and show no enhancements due to shocked ejecta, consistent with a remnant in the Sedov phase. The plasma temperature is relatively soft ($\sim0.2$~keV for the Sedov model), similar to other evolved remnants in the LMC.

\item Using our spectral fit results and the Sedov self-similar solution, we estimated the age of \SNR\ to be $\sim17-27$~kyr, with an initial explosion energy of $(0.09-0.83)\times10^{51}$~erg, and a swept-up interstellar medium mass of 48--162~M$_{\sun}$, further confirming that the SNR is in the Sedov phase.

\item We determined that \source, the source located near the centre of \SNR, can either be a background AGN or a source in the LMC, most likely a LMXB. Because no shocked ejecta emission was detected in the X-ray spectrum, we cannot offer a typing of the progenitor supernova, and the assertion of BF13 that the remnant is the result of a Type Ia event cannot be improved upon.

\end{itemize}

\begin{acknowledgements} We would like to thank our referee Rosa Williams for her valuable comments on how to improve our paper. This research is funded by the Bundesministerium f\"{u}r Wirtschaft und Technologie/Deutsches Zentrum f\"{u}r Luft- und Raumfahrt (BMWi/DLR) through grant FKZ 50 OR 1309. M.S. acknowledges support by the Deutsche Forschungsgemeinschaft through the Emmy Noether Research Grant SA2131/1-1. E.T.W. acknowledges support by the Deutsche Forschungsgemeinschaft through the Research Grant WH-172/1-1. P.\,M. acknowledges support from the BMWi/DLR through grant FKZ 50 OR 1201. \end{acknowledgements} 

\bibliographystyle{aa}

\end{document}